\documentclass[sigconf,authorversion]{acmart}

\AtBeginDocument{%
  \providecommand\BibTeX{{%
    \normalfont B\kern-0.5em{\scshape i\kern-0.25em b}\kern-0.8em\TeX}}}

\copyrightyear{2025}
\acmYear{2025}
\setcopyright{cc}
\setcctype{by}
\acmConference[SIGIR-AP 2025]{Proceedings of the 2025 Annual International ACM SIGIR Conference on Research and Development in Information Retrieval in the Asia Pacific Region}{December 7--10, 2025}{Xi'an, China}
\acmBooktitle{Proceedings of the 2025 Annual International ACM SIGIR Conference on Research and Development in Information Retrieval in the Asia Pacific Region (SIGIR-AP 2025), December 7--10, 2025, Xi'an, China}
\acmDOI{10.1145/3767695.3769478}
\acmISBN{979-8-4007-2218-9/2025/12}

\usepackage{bbm}
\usepackage{booktabs} 
\usepackage{color}
\usepackage{colortbl}
\usepackage{enumitem}
\usepackage{multirow}
\usepackage{makecell}
\usepackage{xcolor}
\usepackage{tabularx}
\usepackage{subcaption}

\setlist[itemize]{leftmargin=*}

\begin{document}

\title{Limitations of Current Evaluation Practices for Conversational Recommender Systems and the Potential of User Simulation}

\author{Nolwenn Bernard}
\authornote{Work done while at the University of Stavanger, Norway}
\affiliation{%
  \institution{TH Köln}
  \city{Cologne}
  \country{Germany}
}
\email{nolwenn.bernard@th-koeln.de}

\author{Krisztian Balog}
\affiliation{%
  \institution{University of Stavanger}
  \city{Stavanger}
  \country{Norway}
}
\email{krisztian.balog@uis.no}

\begin{abstract}
Research and development on conversational recommender systems (CRSs) critically depends on sound and reliable evaluation methodologies. However, the interactive nature of these systems poses significant challenges for automatic evaluation. This paper critically examines current evaluation practices and identifies two key limitations: the over-reliance on static test collections and the inadequacy of existing evaluation metrics. To substantiate this critique, we analyze real user interactions with nine existing CRSs and demonstrate a striking disconnect between self-reported user satisfaction and  performance scores reported in prior literature. 
To address these limitations, this work explores the potential of user simulation to generate dynamic interaction data, offering a departure from static datasets. Furthermore, we propose novel evaluation metrics, based on a general reward/cost framework, designed to better align with real user satisfaction. Our analysis of different simulation approaches provides valuable insights into their effectiveness and reveals promising initial results, showing improved correlation with system rankings compared to human evaluation. While these findings indicate a significant step forward in CRS evaluation, we also identify areas for future research and refinement in both simulation techniques and evaluation metrics.
\end{abstract}



\maketitle

\section{Introduction}

The evaluation of conversational recommender systems (CRSs) is a complex problem, requiring consideration of both system effectiveness and user experience. 
There is a long-standing tradition in recommender systems of adapting evaluation methodologies from information retrieval (IR), a trend seen, for instance, in the evolution from rating prediction to top-k ranking evaluation~\citep{Cremonesi:2010:RecSys,Bellogin:2011:RecSys,Steck:2013:RecSys,Valcarce:2018:RecSys}. However, for CRSs, this established paradigm proves insufficient, as returning relevant recommendations at top ranks does not always guarantee user satisfaction~\citep{Pramod:2022:ESA}. The interactive setting further contributes to the complexity of the evaluation process.  While different evaluation paradigms have been proposed, a standard methodology remains to be established~\citep{Jannach:2023:AIR,Gao:2021:AIOpen}.

In this work, we investigate current practices for automatically evaluating the effectiveness of CRSs, typically performed offline using benchmark conversational datasets (e.g., ReDial~\citep{Li:2018:NIPS} and OpenDialKG~\citep{Moon:2019:ACL}). We identify two main limitations of this approach: (1) reliance on static datasets that overlook the interactive nature of conversations and (2) the inadequacy of traditional metrics to assess user \emph{utility}. To address these fundamental limitations and contribute to establishing a more appropriate evaluation paradigm, our primary contributions are twofold: first, we introduce novel, user-centric metrics grounded in the economic principles of reward and cost during interaction, moving beyond traditional effectiveness measures. Second, we perform an empirical investigation into the use of user simulation for generating synthetic conversations, thereby enabling a more dynamic and automated evaluation framework for CRSs.

Our first contribution, focusing on developing more appropriate user utility metrics, stems from the observation that current evaluation metrics for CRSs often prioritize specific system attributes like recommendation accuracy and fluency, neglecting the ultimate goal of user utility. Moreover, these metrics are frequently borrowed from static settings (e.g., Recall@N from IR, Distinct-n from natural language generation) and are therefore ill-suited for capturing the interactive nature of CRSs. Indeed, we empirically demonstrate that current CRS evaluation based on static test collections and traditional metrics fail to assess the actual utility generated for users. We advocate for a user-centric evaluation approach, emphasizing utility. Consequently, we propose a utility framework, inspired by economic models of human-computer interaction~\citep{Azzopardi:2019:CHI,Azzopardi:2018:bookchapter}, that balances the reward gained and cost incurred for a user interacting with a CRS. Based on this framework, we propose two new metrics and demonstrate their stronger alignment with user satisfaction compared to traditional metrics like Recall@N. Furthermore, we re-examine previously reported CRS performance results~\citep{Wang:2023:EMNLP,Manzoor:2021:RecSys} using our best-performing utility-based metric and find that some conclusions drawn in prior work no longer hold.

Regarding the reliance on static datasets that overlook the interactive nature of conversations, user simulation offers a potential path towards automated CRS evaluation. Its widespread adoption, however, is hindered by questions surrounding the validity of user simulators~\citep{Balog:2024:FnTIR}. Nevertheless, previous work has shown that user simulators can reproduce the ranking of CRSs obtained from offline experiments and user studies~\citep{Wang:2023:EMNLP,Huang:2024:arXiv,Zhang:2020:KDD}, supporting the viability of simulation-based evaluation as an alternative to traditional offline experiments. Building on this promising avenue, our second key contribution involves a comparative analysis of different user simulation approaches. To the best of our knowledge, a direct comparison of user simulators for evaluating CRSs and their alignment with human judgments has not yet been performed. Following the framework proposed in~\citep{Bernard:2024:ICTIR}, we compare the reliability of traditional agenda-based and more recent LLM-based simulators in evaluating nine existing CRSs against CRS rankings in terms of self-reported user satisfaction. Our findings reveal that the LLM-based simulator is more reliable; however, limitations remain in terms of reproducing the relative ranking of a diverse set of CRSs.

To summarize, this paper makes four key contributions that address the limitations of current CRS evaluation.
First, we introduce a general reward-cost framework for measuring user utility in CRS interactions. Second, we propose two novel metrics derived from this framework and demonstrate their positive correlation with self-reported user satisfaction. Third, by re-evaluating the performance of existing CRSs using our best-performing utility metric, we show that certain findings from prior work based on static datasets are no longer supported. Finally, we present a comparative analysis of different user simulator approaches regarding their reliability in evaluating CRSs. 
Overall, this work takes a critical step towards more ecologically valid and user-centric automated CRS evaluation utilizing user simulation. 

\section{Related Work}
\label{sec:related}

This work focuses on the evaluation of conversational recommender systems (CRSs) and the potential of user simulation to alleviate some of the limitations of current offline evaluation practices. In this section, we provide an overview of CRSs types and evaluation methodologies, and user simulation approaches.

\subsection{Conversational Recommender Systems}

We identify four main types of CRS proposed over the years.
First, \emph{rule-based CRSs} rely on a set of predefined rules to generate responses~\citep{Dalton:2018:SIGIR,Habib:2020:CIKM,Thompson:2004:JAIR}. This allows for total control over the system's behavior as well as transparency. However, it also presents disadvantages such as the difficulty of maintaining and updating the rules as the system extends to new domains. 
Second, \emph{retrieval-based CRSs} follow the idea of retrieving the most relevant response from a corpus of predefined responses, including placeholders for recommended items~\citep{Manzoor:2022:InfSys}. This approach is more flexible than the rule-based one, while keeping high quality responses with regards to naturalness and meaningfulness. 
However, having predefined responses limits the diversity and adaptability of the system.
Third, CRSs can be built using deep neural networks, following either a pipeline or end-to-end architecture~\citep{Chen:2019:EMNLP,Zhou:2020:KDD,Wang:2022:KDD}. 
These systems are more flexible to unknown contexts and easier to adapt to new domains, given sufficient training data.
However, using deep neural networks reduces the transparency of the system and its controllability due to the black-box nature of these models. 
Fourth, the most recent type of CRS relies on large language models (LLMs)~\citep{Friedman:2023:arXiv,Sun:2024:arXiv,He:2023:CIKM}. LLMs have shown great potential in generating human-like responses which is critical in a conversational setting. The recommendation task may be performed by the LLM itself based on its internal knowledge or in combination with a recommender system.
LLM-based CRSs require less even no data for training considering the pre-trained models available. Hence, they appear as data-efficient and accessible solutions for conversational recommendation. 
However, similarly to neural-based CRSs, their transparency and controllability is limited.  

\subsection{Evaluation Methodologies} 

Traditionally, the main methodologies to evaluate CRSs are:  online, user studies, and offline evaluation~\citep{Balog:2024:FnTIR}.
\emph{Online evaluation} requires a real-world setting where real users interact with a CRS, thus, providing arguably the most reliable evaluation. In practice, it often corresponds to A/B tests where the performances of different versions of a CRS are compared with regards to a set of performance metrics. Such evaluation can be difficult to conduct, especially in an academic setting, because it requires access to a sufficient number of real users and different versions of the CRS to compare. 
\emph{User studies} involve human participants that are not necessarily actual users of the system. These studies, conducted in a laboratory or crowdsourcing setting, allow for the assessment of a CRS with regards to different characteristics at different granularity levels~\citep{Jannach:2023:AIR}. For example, \citet{Manzoor:2024:UMAP} and \citet{Manzoor:2021:RecSys} study the meaningfulness of system responses, while, \citet{Li:2018:NIPS} assess the quality of system response at the conversation level.
Finally, \emph{offline evaluation} does not rely on human participants 
and is commonly used to evaluate algorithms and models using automatic metrics such as Recall and BLEU.
Recently, LLMs are being employed as an alternative to human judges for text evaluation~\citep{Hashemi:2024:ACL,Chiang:2023:ACL}. This makes it possible to automatically evaluate more ``subjective'' aspects, such as the quality of explanations~\citep{Huang:2024:arXiv} and conciseness~\citep{Hashemi:2024:ACL}.

\emph{Simulation-based evaluation}, which is gaining attention in the field, bridges the gap between online/user studies and offline evaluation. Indeed, user simulators aim to mimic the behavior of human users, therefore, simulated users can substitute real users to perform the evaluation of CRSs in a controlled and interactive manner~\citep{Balog:2024:FnTIR}.
There are two main strategies to using simulation: (1) the user simulator interacts with the CRS and provides direct feedback on the system's performances~\citep{Sun:2021:SIGIR}, and (2) the user simulator interacts with the CRS to generate dialogues that are separately evaluated~\citep{Wang:2023:EMNLP,Huang:2024:arXiv,Zhang:2020:KDD}.
Despite recent advances, simulation-based evaluation is still dependent on benchmark datasets. For example, \citet{Wang:2023:EMNLP} generate synthetic dialogues based on excerpts of dialogues from ReDial~\citep{Li:2018:NIPS} or OpenDialKG~\citep{Moon:2019:ACL} and stop either when a ground truth item is recommended or a predetermined number of turns is reached, leading to potentially unnatural endings. 

\begin{figure*}
    \centering
    \includegraphics[width=\textwidth,keepaspectratio]{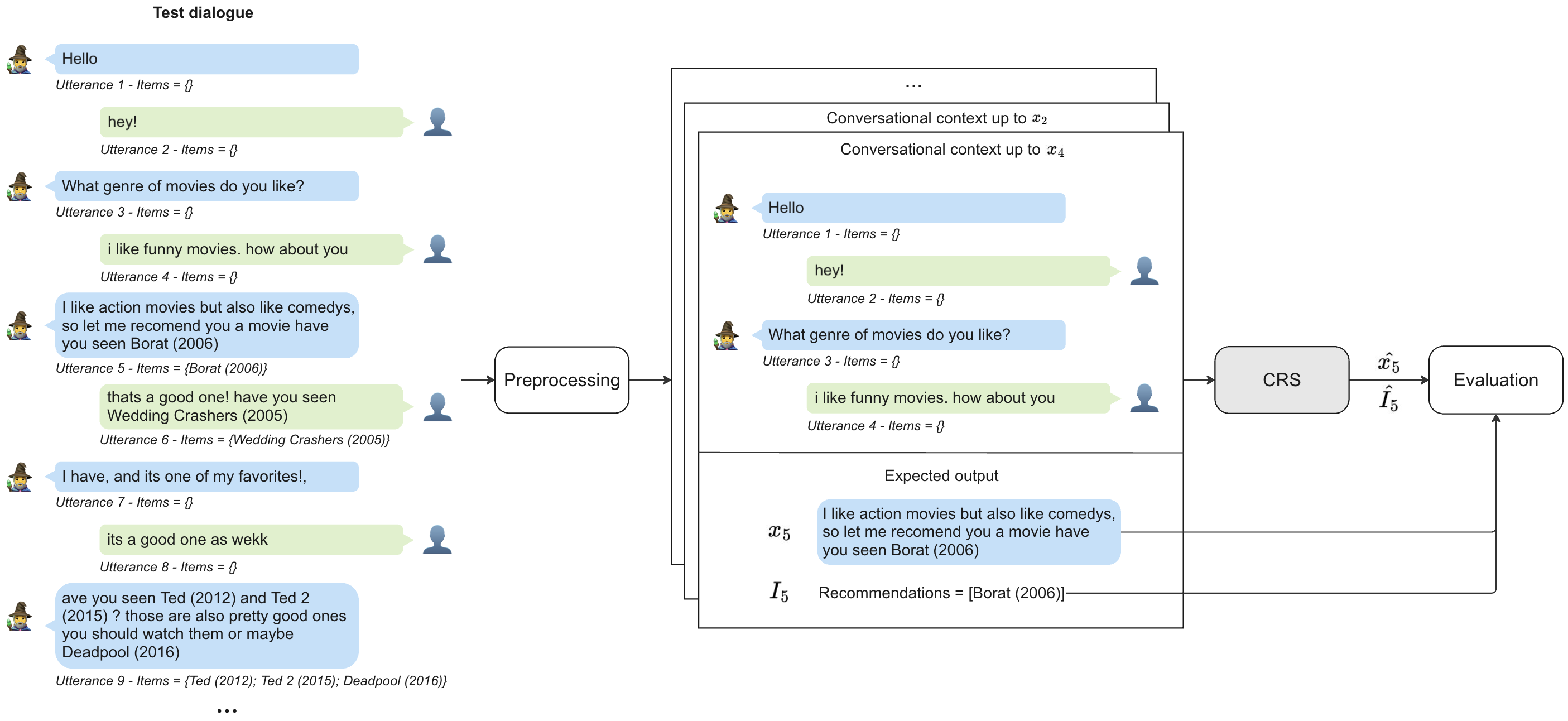}
    \caption{Illustration of the current CRS evaluation practice, based on static conversational datasets.}
    \label{fig:eval-practice}
\end{figure*}

\subsection{User Simulation Approaches}

User simulators share architectural similarities with CRSs, given that both are intelligent agents that need to generate utterances, which involves the tasks of natural language understanding, dialogue management, and natural language generation.
We identify three types of user simulator: agenda-based, deep neural-based, and LLM-based.
\emph{Agenda-based user simulators} determine the next action based on an agenda built from a goal and that is updated based on the context of the conversation~\citep{Schatzmann:2007:NAACL}. A Markovian state representation is assumed, and the allowed state transitions are specified by an interaction model. This design allows for transparency of the simulated users' behavior. However, the creation of such user simulators requires a certain knowledge of the domain and task, especially for the creation of the interaction model.
To address this issue, \emph{data-driven user simulators} have been proposed, which often rely on deep neural networks to learn  user behavior from data directly~\citep{Lin:2021:SIGDIAL}. This type of user simulators often abstracts the response generation process, making it less interpretable. Similarly to neural-based CRSs, the transparency of these simulators is limited and they require sufficient data for training.
Finally, the rise of LLMs has also impacted the creation of user simulators. Indeed, recent works have proposed to build user simulators based on LLMs for conversational information access and task-oriented dialogue systems~\citep{Abbasiantaeb:2024:WSDM,Huang:2024:arXiv,Wang:2023:EMNLP,Lin:2022:SIGDIAL,Sekulic:2024:SCICHAT,Yoon:2024:NAACL}. For example, \citet{Abbasiantaeb:2024:WSDM} use LLM-based user simulators for data generation in the context of conversational question answering, while  \citet{Huang:2024:arXiv} and \citet{Wang:2023:EMNLP} use LLM-based user simulators for CRS evaluation.

\section{Current Evaluation Practices and their Limitations}
\label{sec:limits}

This section first details the predominant approach to automatic CRS evaluation, and then identifies two key limitations: the reliance on offline datasets and the choice of metrics.

\subsection{Problem Statement}

In this work, we focus on the evaluation of CRSs. The objective of a CRS is to assist users in satisfying recommendation-related goals through a multi-turn dialogue~\citep{Jannach:2021:CSUR}. The CRS and the user are active dialogue participants that take turns in issuing utterances $x_i$. Therefore, a dialogue is defined as a sequence of utterances: $d = \{x_1, x_2, \ldots, x_n\}$. We note that the utterances can be annotated with semantic information, such as dialogue acts~\citep{Jurafsky:2023:Book}, as well as metadata. The metadata typically includes items $I$ that are recommended by the CRS or mentioned by the user.

\subsection{Current Practice}

Offline evaluation of CRSs typically involves three stages, leveraging annotated conversational datasets, illustrated in Fig.~\ref{fig:eval-practice}. First, dialogues from the test portion of the dataset (hereinafter: test dialogues) are preprocessed and/or augmented to derive inputs and expected outputs for the CRS.  
The input corresponds to the conversational context up to utterance $x_t$ (issued by the user) along with optional metadata (e.g., entity mentioned). 
The expected output corresponds to the next utterance $x_{t+1}$ and associated recommendations $I_{t+1}$ (which may be empty) in the test dialogue. Second, the CRS generates the next utterance $\hat{x}_{t+1}$ and the associated recommendations $\hat{I}_{t+1}$ given the input. Third, evaluation metrics are computed on the \emph{turn level}, and then aggregated (micro-averaged or macro-averaged). 
Traditionally, the metrics assess either \emph{task performance} in terms of recommendation accuracy, by comparing $\hat{I}_{t+1}$ with the ground truth target items $I_{t+1}$, or the quality of the generated utterance $\hat{x}_{t+1}$ either with respect to the expected output utterance $x_{t+1}$ or on its own account (e.g., fluency). This process is repeated and the results are aggregated over all test dialogues.

\subsection{Limitation \#1: Reliance on Offline Datasets}

The previous description of the traditional offline evaluation process highlights the static nature of the datasets used. By definition, they neglect the interactive nature of conversational systems~\citep{Balog:2024:FnTIR}.
The recommended items in $I_{t+1}$, which are often annotated after the data collection, are considered as ground truth. 
We observe that typically, $I_{t+1}$ comprises a single item, for example, it represents more than 80\% of the cases in the ReDial dataset~\citep{Li:2018:NIPS}. This is unrealistic, as a user is unlikely to look for a specific item when interacting with a CRS, but rather for recommendations that satisfy their current preferences, consequently the effectiveness of the CRSs may be underestimated~\citep{Vlachou:2024:arXiv}.
Moreover, \citet{He:2023:CIKM} notice that items in $I$ may be repeated during the conversation, i.e., ground truth items comprise items that were already mentioned in the conversation, which can lead to overestimating the effectiveness of CRSs.
Furthermore, the lack of diversity in dialogue patterns within offline datasets has also been noted~\citep{Jannach:2023:AIR}.
To mitigate these limitations, in this work, we use dialogues collected with real users and synthetic dialogues generated with user simulators that are guided by an overall goal~\citep{Schatzmann:2007:NAACL} rather than specific target item(s).

\subsection{Limitation \#2: Evaluation Metrics}

Automatic metrics used in offline evaluation often focus on specific system aspects, such as recommendation accuracy or linguistic diversity, rather than overall user utility. For instance, Recall@N is a commonly used metric to evaluate recommendation accuracy (see, e.g.,~\cite{Chen:2019:EMNLP,Wang:2022:KDD,Zhou:2020:KDD,Wang:2023:EMNLP,Huang:2024:arXiv}), typically computed with $N$ set to 1, 10, 25, and 50. Our review of the literature does not provide a clear rationale for the choice of large $N$ values ($N>10$). We argue that in a conversational setting, where utterances should remain focused and concise, users expect targeted and manageable suggestions rather than a large batch of items that are likely to overwhelm them. 
However, to the best of our knowledge, this point has not been previously discussed in the literature, and these values for $N$ have been widely adopted in the field.
We hypothesize that the prevalent use of larger $N$ values might stem from the common assumption of a single relevant ground truth item in the evaluation dataset (as discussed in Limitation \#1). In such scenarios, lower $N$ values could indeed lead to metrics that are too insensitive to effectively differentiate between recommendation algorithms. However, this methodological choice carries the significant risk of the evaluation metric becoming increasingly detached from real-world user experience and satisfaction.
Indeed, in recent work, \citet{Bernard:2025:WSDM} found a negligible correlation between Recall@10 and self-reported user satisfaction from real users, suggesting that Recall@N may not be  a reliable proxy for user utility. However, they do not propose alternative metrics. 
Similarly, some studies report distinct-n to assess the linguistic diversity of the system without considering the conversational context~\citep{Chen:2019:EMNLP,Zhou:2020:KDD,Wang:2022:KDD}. However, this approach, in our view, does not accurately reflect a CRS's ability to assist users in finding relevant items.
Furthermore, some of the commonly used metrics were originally designed for static tasks, such as recommendation or language generation, and thus often overlook the interactive nature of conversational systems.
In this work, we propose alternative metrics that evaluate CRSs with regards to user utility, considering the conversational nature of the task and lifting the assumption of the user looking for a single specific target items.

\section{Evaluation Metrics}
\label{sec:metrics}

The ultimate goal of a conversational recommender system is to generate utility for its users~\citep{Jannach:2023:AIR}, which should ideally manifest as high user satisfaction.
However, as demonstrated in~\citep{Bernard:2025:WSDM}, and also confirmed by our experiments in Section~\ref{sec:expeval}, the widely used Recall@N measure fails to adequately capture this utility. 
This leads to our central research question for this section: \emph{Can we develop evaluation metrics for CRSs focused on user utility that show a greater correlation with overall end-user satisfaction?}

To address this question, we first propose a formal framework to provide a more direct measure of user utility by considering both the effort expended by the user (i.e., cost) during their interactions with the CRS and the benefit (i.e., reward) gained from the recommendations made by the system (Section~\ref{sec:metrics:fw}).
We show how this framework can encapsulate existing evaluation measures (Section~\ref{sec:metrics:existing}) and then propose novel evaluation metrics that aim to provide better approximations of costs and rewards (Section~\ref{sec:metrics:novel}).

\subsection{A General Reward/Cost Framework}
\label{sec:metrics:fw}

Evaluating the effectiveness of a CRS requires considering not only the benefits users derive but also the costs associated with the interaction.
To this end, we propose a general utility framework for measuring the effectiveness of a CRS inspired by economic models of human-computer interaction~\citep{Azzopardi:2019:CHI,Azzopardi:2018:bookchapter}.
Our measure of utility is based on the balance between the \emph{reward} $R$ the user is gaining from interacting with the system and the overall \emph{cost} $C$ incurred throughout the conversation. 

We assume that there are multiple reward factors $r_i$, each with its associated weight $\alpha_i$, that can contribute to the overall reward: $R=\sum_i \alpha_i r_i$. For example, the CRS making relevant recommendations, exposing the user to diverse items, or enabling serendipitous discoveries are all positive outcomes where the user gains from interacting with the system.
Similarly, we write cost as a weighted linear combination of various cost factors: $C=\sum_j \beta_j c_j$.
Examples of cost factors include the interaction length (total number of turns), the cognitive load associated with understanding the system's questions and formulating responses, and the time the user has to wait for the system to respond.

The utility of the conversational recommendation process is then defined as the ratio of the total reward to the total cost:
\begin{equation}
    \mathrm{utility} = \frac{R}{C} ~. \label{eq:utility}
\end{equation}
A higher utility score indicates a more effective CRS, where the user gains significant reward with relatively low interaction costs.
Utility, as defined in Eq.~\eqref{eq:utility} for a single conversation, is then aggregated into a final system-level metric by averaging its value across a set of evaluation conversations.

We note that this framework offers a general approach applicable to various interactive systems. However, our primary focus in this work lies specifically in its instantiation for the task of conversational recommendations.
In the subsequent sections, we will discuss relatively simple instantiations of this framework, considering a single reward factor and a single cost factor at a time. This simplification allows us to demonstrate the potential of measuring effectiveness from an economic perspective, without the need to define specific weight coefficients ($\alpha_i$ and $\beta_j$). The development and analysis of more advanced metrics incorporating multiple weighted factors and normalization techniques are left for future work.

\subsection{Existing Measures}
\label{sec:metrics:existing}

We show how two of the most commonly used CRS effectiveness measures can be instantiated in our reward/cost framework. \\

\noindent
\emph{\textbf{Recall@N}.}
This measure is only computed for turns where a recommendation has been made. Then, the turn-level recall values are aggregated on the dataset level,\footnote{This is not explicitly stated in any of the papers reporting on Recall@N and is based on our understanding of the implementation in CRSLab~\citep{Zhou:2021:IJCNLP}, a general toolkit to develop and evaluate CRSs (see, e.g.,~\citep{Zhou:2022:WSDM,Wang:2022:KDD}).} corresponding to micro-averaging. As such, it does not directly measure the effectiveness of in a given dialogue, which explains why it does not correlates with user satisfaction. All other measures we discuss in this paper perform macro-averaging instead, by first evaluating conversations and then averaging over all conversations in the test dataset.

In terms of the Reward/Cost framework, the total cost, across all turns $t$ in all test dialogues $d$, is the total number of item recommendations made. The total reward is the number of recommended items that match the ground truth target items. Formally:
\begin{equation}
    Recall@N = \frac{1}{\sum_d |d|} \frac{\sum_d \sum_t |I_t \cap \hat{I}_t|}{\sum_d \sum_t N \mathbbm{1}_{rec}(t)} ~,
\end{equation}
where $|d|$ is the number of turns in a dialogue and $\mathbbm{1}_{rec}(t)$ is an indicator function that is 1 if a recommendation was made in turn $t$ and 0 otherwise. \\

\noindent
\emph{\textbf{Success Rate (SR)}.}
A simple and intuitive way to assess a CRS is based on the ratio of ``successful'' dialogues in the test dataset $D$. Success is usually determined by checking if any recommendation matches any of the ground truth target items~\citep{Lei:2020:WSDM,Sun:2018:SIGIR,Xu:2021:WSDM,Yu:2024:TKDE}. 
That is, the reward for a single conversation is a binary value indicating whether there was a successful recommendation or not: $R=\mathbbm{1}_{succ}(d)$. The cost is not considered, i.e., it's a constant for all dialogues: $C=1$. Putting these choices together, success rate on a set of dialogues $D$ is measured as:
\begin{equation}
    SR = \frac{\sum_d \mathbbm{1}_{succ}(d)}{|D|} ~. \label{eq:sr}
\end{equation}

\subsection{Novel Measures}
\label{sec:metrics:novel}

Current CRS evaluation is limited by its dependence on ground truth target items (typically a single item). Additionally, these metrics largely ignore the cost incurred by the user during the interaction, or only consider it selectively during recommendation turns. We propose two novel user-centric metrics that consider the underlying intent of the user utterances. We operationalize this intent by leveraging the concept of a \emph{dialogue act}, which is a structured representation of the of the communicative goal behind an utterance~\citep{Jurafsky:2023:Book}. \\

\noindent
\emph{\textbf{Successful Recommendation Round Ratio (SRRR)}.}
Success rate is a binary metric that does not account for multiple recommendation rounds in a dialogue, which can impact the cognitive load for user (i.e., more recommendations to assess). 
We introduce the notion of a \emph{recommendation round} in a dialogue: a $round$ starts when the intent of the CRS is \texttt{Recommend} and finishes when the user makes a decision to either \texttt{Accept} or \texttt{Reject} the recommendation. 
Each recommendation round has a fixed cost $C=|rounds|$. Successful recommendation rounds (ending with an acceptance) yield a reward of 1, otherwise 0: $R=\mathbbm{1}_{acc}(round)$. Putting these choices together:
\begin{equation}
    SRRR = \frac{1}{|D|} \sum_d \frac{\sum_{round} \mathbbm{1}_{acc}(round)}{|rounds|} .
\end{equation}

\noindent
\emph{\textbf{Reward-per-Dialogue-Length (RDL)}.}
Notice that all the previous measures either do not consider cost at all or only consider the turns where a recommendation has been made. Instead, we argue to consider all conversation turns as part of the effort expended by the user. A simple way of doing that is setting the cost to the total number of turns in the dialogue: $C=|d|$.
The reward is the total number of recommendations accepted by the user $R=|I_{acc}|$. Then, the reward-per-dialogue-length on a set of dialogues is:
\begin{equation}
    RDL = \frac{1}{|D|} \sum_d  \frac{|I_{acc}|}{|d|} ~.
\end{equation}

\section{Experimental Evaluation}
\label{sec:expeval}

This section presents an experimental comparison of various evaluation metrics, focusing on their correlation with self-reported user satisfaction.
Furthermore, we examine the performance of the different CRSs when interacting with real users.

\subsection{Experimental Setup}

\begin{table}
    \centering
    \caption{Summary of dialogue corpora in CRSArena-Dial~\citep{Bernard:2025:WSDM}.}
    \vspace*{-0.7\baselineskip}
    \label{tab:datasets}
    \begin{tabular}{lc}
        \toprule
        \textbf{CRS} & \textbf{\#dialogues} \\
         \midrule
        CRB-CRS\_ReDial & 60  \\
        KBRD\_OpenDialKG & 59 \\
        KBRD\_ReDial & 61  \\
        BARCOR\_OpenDialKG & 55 \\
        BARCOR\_ReDial & 46  \\
        UniCRS\_OpenDialKG & 42  \\
        UniCRS\_ReDial & 48  \\
        ChatCRS\_OpenDialKG & 44  \\
        ChatCRS\_ReDial & 52  \\ 
        \midrule
        Total & 467  \\ 
        \bottomrule
    \end{tabular}
\end{table}

For our experiments, we utilize conversations from the CRSArena-Dial dataset~\citep{Bernard:2025:WSDM}. This dataset comprises 467 conversations real users conducted with nine different CRSs; see Table~\ref{tab:datasets} for a breakdown per CRS. Additionally, this dataset contains self-reported user satisfaction (i.e., satisfied or frustrated), and direct and free-form feedback from users with regards to their experience with the CRSs.

In order to compute the different utility metrics, we automatically annotate the utterances with dialogue acts.
For example, the utterance ``Have you seen Step Brothers or Walk Hard: The Dewey Cox Story'' can be annotated with the following dialogue act: \texttt{Reco\-mmend(TITLE=`Step Brothers', TITLE=`Walk Hard: The Dewey Cox Story')}. The annotation is performed following a few-shot prompting approach with the Mistral NeMo\footnote{\url{https://mistral.ai/news/mistral-nemo/}} large language model; the intent schema is taken from the IARD dataset~\citep{Cai:2020:UMAP}.

Given that we do not have ground truth item annotations, we take Recall@1 values from \citep{Wang:2022:arXiv} and measure whether a recommendation was successful or not ($R=\mathbbm{1}_{succ}(d)$ in Eq.~\ref{eq:sr}) based on the presence of dialogue acts with the intent \texttt{Accept} in user utterances.

\subsection{Conversational Recommender Systems}

We consider nine CRSs based on five different architectures, tailored for the movie recommendation domain.\footnote{We use their implementation from CRSArena at: \url{https://huggingface.co/spaces/iai-group/CRSArena/tree/main}.} The CRSs are trained and/or leverage external knowledge from either a subset of OpenDialKG~\citep{Moon:2019:ACL} or ReDial~\citep{Li:2018:NIPS}. The selection of these CRSs is guided by their public availability and use in previous studies. While we acknowledge limitations such as the emphasis on publicly available English-speaking CRSs, this selection represents a broad range of systems based on existing research. 

\begin{itemize}
    \item \textbf{CRB-CRS}~\citep{Manzoor:2022:InfSys} comprises two main components: (1) a retrieval and ranking component and (2) a recommendation and metadata integration component. The first is responsible for finding the best pre-defined response candidate from a pool based on the last user utterance and conversation history. The second adapts the previously retrieved response candidate by replacing placeholders with the recommended items and other relevant information. 
    \item \textbf{KBRD}~\citep{Chen:2019:EMNLP} uses knowledge propagation to integrate recommendation and dialogue systems for mutual enhancement. The recommender system takes advantage of the dialogue context to provide better recommendations, and the dialogue system uses the information about the recommended items and user model to generate more consistent responses. KBRD uses an external knowledge graph to enhance the representation of entities in the dialogue context and recommended items. The propagation of knowledge-enhanced representations helps to find connected items to recommend and to generate more informative responses.
    \item \textbf{BARCOR}~\citep{Wang:2022:arXiv} performs the recommendation and dialogue generation tasks using a single model, based on the pre-trained language model BART~\citep{Lewis:2020:ACL}. Specifically, it uses the encoder for the recommendation task and the decoder to generate responses. BART is fine-tuned with examples where a specific token is used to mask the recommended items.
    \item \textbf{UniCRS}~\citep{Wang:2022:KDD} uses the prompt learning paradigm to fulfill the tasks of recommendation and conversation in an unified manner. UniCRS relies on knowledge-enhanced prompts which include fused knowledge representation, task specific tokens, and dialogue context, in addition to a response template for the recommendation task. These prompts are then processed by the pre-trained language model DialoGPT~\citep{Zhang:2020:ACL}.
    \item \textbf{ChatCRS}~\citep{Wang:2023:EMNLP} combines ChatGPT and a recommendation model to perform conversational recommendation. The recommendation model computes the similarity between the embedded conversational history and embedded candidate items. Item embeddings are created using the model \emph{text-embedding-ada-002}~\citep{Neelakantan:2022:arXiv} from item attributes such as title, genre, and actor for movies. While ChatCRS can directly recommend items, the use of candidate items limits hallucinations and recommendation of items outside of the evaluation dataset.
\end{itemize}

\subsection{Results}
\label{sec:expeval:results}

\begin{table*}
    \centering
    \small
    \caption{Evaluation of CRSs with respect to self-reported user satisfaction (taken from \citep{Bernard:2025:WSDM}), which is taken as the ground truth (top row). Correlations are measured using Kendall's $\tau$.
    The retrieval-based CRS is highlighted in \textbf{\textcolor{blue}{bold blue}} and the CRSs with the best architecture according to previous work~\citep{Huang:2024:arXiv,Manzoor:2024:UMAP,Wang:2023:EMNLP} are highlighted in \textbf{\textcolor{orange}{bold orange}}.
    }
    \vspace*{-0.7\baselineskip}
    \label{tab:utilities-rankings}
    \begin{tabular}{p{2.2cm}cp{0.73\textwidth}}
        \toprule
        \textbf{Metric} & \textbf{Kendall's $\tau$} & \textbf{Ranking} \\ 
        \toprule
        \rowcolor{gray!15}
        \textbf{User satisfaction} & -- & \textbf{\textcolor{orange}{ChatCRS\_OpenDialKG}} (0.523) $\succ$ \textbf{\textcolor{orange}{ChatCRS\_ReDial}} (0.453) $\succ$ BARCOR\_ReDial (0.298) $\succ$ BARCOR\_OpenDialKG (0.145) $\succ$ UniCRS\_ReDial (0.102) $\succ$ KBRD\_ReDial (0.081) $\succ$ \textbf{\textcolor{blue}{CRB-CRS\_ReDial}} (0.079) $\succ$ UniCRS\_OpenDialKG (0.048) $\succ$ KBRD\_OpenDialKG (0.017) \\
        \toprule        
        \textbf{Recall@1}\textsuperscript{\textdagger} & 0.07 & BARCOR\_OpenDialKG (0.312) $\succ$ \textbf{\textcolor{orange}{ChatCRS\_OpenDialKG}} (0.310) $\succ$ UniCRS\_OpenDialKG (0.308) $\succ$ KBRD\_OpenDialKG (0.231) $\succ$ UniCRS\_ReDial (0.050) $\succ$ \textbf{\textcolor{orange}{ChatCRS\_ReDial}} (0.037) $\succ$ BARCOR\_ReDial (0.031) $\succ$ KBRD\_ReDial (0.028) \\
        \midrule        
        \textbf{Success rate} & 0.67 & \textbf{\textcolor{orange}{ChatCRS\_OpenDialKG}} (0.114) $\succ$ \textbf{\textcolor{blue}{CRB-CRS\_ReDial}} (0.111) $\succ$ \textbf{\textcolor{orange}{ChatCRS\_ReDial}} (0.094) $\succ$ BARCOR\_OpenDialKG (0.091) $\succ$ BARCOR\_ReDial (0.085) $\succ$ UniCRS\_ReDial (0.082) $\succ$ KBRD\_ReDial (0.048) $\succ$ UniCRS\_OpenDialKG (0.024) $\succ$ KBRD\_OpenDialKG (0.017) \\ 
        \midrule                
        \textbf{Successful rec.\phantom{xx} round ratio} & 0.32 & \textbf{\textcolor{blue}{CRB-CRS\_ReDial}} (0.077) $\succ$ \textbf{\textcolor{orange}{ChatCRS\_ReDial}} (0.031) $\succ$ BARCOR\_OpenDialKG (0.030) $\succ$ BARCOR\_ReDial (0.023) $\succ$ UniCRS\_ReDial (0.021) $\succ$ \textbf{\textcolor{orange}{ChatCRS\_OpenDialKG}} (0.018) $\succ$ KBRD\_OpenDialKG (0) $\succ$ KBRD\_ReDial (0) $\succ$ UniCRS\_OpenDialKG (0) \\ 
        \midrule                
        \textbf{Reward-per-Dialogue-Length} & 0.78 & \textbf{\textcolor{orange}{ChatCRS\_OpenDialKG}} (0.037) $\succ$ \textbf{\textcolor{orange}{ChatCRS\_ReDial}} (0.025) $\succ$ BARCOR\_ReDial (0.024) $\succ$ \textbf{\textcolor{blue}{CRB-CRS\_ReDial}} (0.023) $\succ$ UniCRS\_ReDial (0.019) $\succ$ BARCOR\_OpenDialKG (0.018) $\succ$ KBRD\_ReDial (0.008) $\succ$ UniCRS\_OpenDialKG (0.006) $\succ$ KBRD\_OpenDialKG (0.004) \\ 
        \bottomrule
        \multicolumn{3}{l}{\textdagger~CRB-CRS\_ReDial is not included as it is not evaluated in~\citep{Wang:2023:EMNLP}.}
    \end{tabular}
\end{table*}

We evaluate the different utility metrics based on their correlation with self-reported user satisfaction. That is, we compute the Kendall's $\tau$ correlation between the rankings of CRSs obtained with each evaluation metric presented in Section~\ref{sec:metrics} and the ranking based on self-reported user satisfaction. The results, reported in Table~\ref{tab:utilities-rankings}, show that \emph{Success Rate} (SR), \emph{Successful Recommendation Round Ratio} (SRRR), and \emph{Reward-per-Dialogue-Length} (RDL) are positively correlated with self-reported user satisfaction, with Kendall's $\tau$ values of 0.67, 0.32, and 0.78, respectively. It suggests that, in our case, Reward-per-Dialogue-Length is the best proxy for user satisfaction among the utility metrics considered. Interestingly, we observe that Success Rate has a higher correlation with user satisfaction than Successful Recommendation Round Ratio, despite the latter being a more fine-grained measure. This may be explained by that SRRR only considers recommendation rounds as part of the cost and the identification of recommendation rounds might be prone to errors due to imperfections of automatic dialogue act annotation.

\subsection{Analysis}

Having established Reward-per-Dialogue-Length (RDL) as the best proxy for user satisfaction, we further analyze the relative rankings of the CRSs based on self-reported user satisfaction and RDL, and evaluate the validity of prior claims made using Recall@N.
We make the following observations:
\begin{itemize}
    \item We observe that a CRS with the same architecture trained with ReDial is generally better than the one trained with OpenDialKG according to both user satisfaction and RDL, with the exception of \emph{ChatCRS}. It suggests that ReDial is a more suitable dataset for training a CRS in the movie recommendation domain. This contrasts with the ranking based on Recall@1 where the CRSs trained with OpenDialKG outperform those trained with ReDial. 
    \item User satisfaction is generally low, with a maximal satisfaction of 0.523 for \emph{ChatCRS\_OpenDialKG}, which is consistent with the low RDL scores observed. This shows a clear inability of the CRSs to efficiently assist users in fulfilling their needs and highlights the need for future work.
    \item The results show that \emph{ChatCRS} outperforms the other CRSs in terms of user satisfaction and RDL. This is consistent with conclusions from previous work~\citep{Huang:2024:arXiv,Manzoor:2024:UMAP,Wang:2023:EMNLP}.
    \item We observe that \emph{CRB-CRS\_ReDial}, which uses retrieval-based responses, is in the top half of the rankings for RDL. In other words, it outperforms some neural CRSs using generation-based responses, such as \emph{UniCRS\_ReDial} and \emph{KBRD\_ReDial}. However, \emph{CRB-CRS\_Re\-Dial} is in the lowest half for self-reported user satisfaction. Moreover, in this ranking, \emph{KBRD\_ReDial} is better than \emph{CRB-CRS\_Re\-Dial}, it contrasts with the findings from the user study in~\citep{Manzoor:2021:RecSys}, which showed that \emph{CRB-CRS} outperforms \emph{KBRD} in terms of response's meaningfulness.
    Direct feedback from CRSArena-Dial provides some insights on the difference between the rank regarding RDL and user satisfaction, that is, some users reported that \emph{CRB-CRS\_ReDial} was able to provide some relevant recommendations, but other factors such as the response time, stubbornness, and misunderstandings led to a feeling of frustration.
\end{itemize}

\section{User Simulation-based Evaluation}

A significant bottleneck in the evaluation of CRSs lies in the reliance on human users to generate interaction data. Collecting and analyzing conversations with real users is a time-consuming and resource-intensive process, hindering the scalability of evaluation efforts, particularly when comparing numerous systems. This motivates our research question: \emph{Can user simulation offer a viable solution to automate the evaluation process, and if so, to what extent can simulations reliably predict the performance of CRSs with real users?} This section delves into the potential of user simulation as a means of automating the collection of interaction data for CRS evaluation, exploring the capabilities and limitations of two fundamentally different simulation approaches in mirroring real user conversations.

\subsection{User Simulators}

We consider two user simulators for movie recommendation:

\begin{itemize}
    \item \textbf{ABUS} is an agenda-based user simulator with a traditional modular architecture comprising modules for natural language understanding and generation and a dialogue manager. In this case, natural language understanding and generation are  performed with Mistral NeMo using a few-shot prompting approach, where the examples are taken from the IARD dataset~\citep{Cai:2020:UMAP}. To create and update the agenda, the dialogue manager relies on an interaction model built using a combination of handcrafted and data-driven rules. The data-driven rules are derived from a subset of the IARD dataset~\citep{Cai:2020:UMAP}.
    \item \textbf{LLM-US} is a simulator relying on the Mistral NeMo LLM. In practice, upon reception of an utterance, \emph{LLM-US} first decides to either continue the conversation or abort it because it does not progress towards the user goal. If the conversation is not aborted, \emph{LLM-US} generates the next utterance using a zero-shot prompting approach using a prompt inspired by \citet{Terragni:2023:arXiv}. We choose zero-shot prompting mainly for its simplicity, however, we acknowledge that other approaches, such as few-shot prompting and specific prompts per action, may be considered for future work.
\end{itemize}
We base our implementation on the UserSimCRS toolkit~\citep{Afzali:2023:WSDM}. 
The original implementation of \emph{ABUS} is adapted to support LLM-based natural language understanding and generation. 
\emph{LLM-US} is added as a new user simulator, along with a wrapper to interact with the CRSs. Note that both user simulators use the LLM Mistral NeMo 12b\footnote{Information related to the model is available at: \url{https://ollama.com/library/mistral-nemo:12b/blobs/b559938ab7a0}} hosted on an Ollama server.\footnote{\url{https://ollama.com}} For the experiments, we keep the default model options except for the temperature which is set to 0 and deactivate the stream option. This model is chosen for its open-source availability and demonstrated performances in natural language generation tasks. Moreover, we release the different prompts used by both simulators.\footnote{\url{https://github.com/iai-group/recsys25-crseval}}

\subsection{Synthetic Conversation Generation}

For the simulation-based evaluation of the CRSs, we generate 200 synthetic dialogues for each (user simulator, CRS) pair, except for \emph{ABUS-ChatCRS\_OpenDialKG} where the conversation generation failed due to recursion errors. To generate a conversation, the user simulator is initialized with a goal prior to the start of the conversation and then interacts with the CRS with the objective of completing that goal. The goal comprises a set of constraints regarding the movies it is interested in, such as the genre and the year of release, and a set of requests corresponding to information it wants to get on the recommended movies, like its plot and director. An example of a goal is:
$$\text{constraints}=\begin{bmatrix}\text{year} = 2016\\\text{actor} = \text{Nicole Kidman}\end{bmatrix} \hspace{0.2cm}
\text{requests}=\begin{bmatrix}\text{plot}\\\text{director}\\\text{rating}\end{bmatrix}$$
One advantage of this approach is that it allows for the generation of diverse conversations, as there are many possible combinations of constraints and requests. 
Furthermore, the user simulator is not looking for any specific target movie, which might get leaked during the conversation. Instead, it considers that there are multiple valid movies satisfying the constraints that may be recommended.
This contrasts with the approach taken by~\citep{Wang:2023:EMNLP}, where a list of target items is directly provided to the user simulator, a scenario we argue is often unrealistic. Indeed, in typical real-world interactions with a CRS, a user does not have a specific movie in mind but is rather seeking recommendations that satisfy their current preferences.

Following the same approach as in Section~\ref{sec:expeval}, we automatically annotate utterances with dialogue acts.

\subsection{Results}

We evaluate CRSs in terms of Reward-per-Dialogue-Length (RDL) and report the results in Table~\ref{tab:utility-diff}. 
Following the framework proposed by~\citet{Bernard:2024:ICTIR}, we examine the reliability of the user simulators by comparing the results obtained with simulated users (Table~\ref{tab:utility-diff}) against those obtained with real users (Table~\ref{tab:utilities-rankings}) serving as the reference. The comparison focuses on two aspects: (1) the ability to reproduce the relative ranking of CRSs and (2) the approximation of absolute performance measurements. 

For the first aspect, we compute the Kendall's $\tau$ correlation between the rankings of CRSs obtained with the simulators and the one based on self-reported user satisfaction. We find correlations of 0.366 for \emph{LLM-US} and 0.143 for \emph{ABUS}, indicating an advantage for \emph{LLM-US}. This is specifically observed for pairwise comparisons of CRSs with the same underlying architecture, where \emph{LLM-US} ranks them in accordance with user satisfaction, except for \emph{UniCRS\_OpenDialKG} and \emph{UniCRS\_ReDial}. However, while the correlations are positive, they are weak, suggesting that the simulators are not yet robust enough to assess a diverse set of CRSs. 

Regarding the second aspect, we report the absolute differences between RDL obtained from the conversations with real users and with simulated users in Table~\ref{tab:utility-diff}. The results show that \emph{LLM-US} is better than \emph{ABUS} at approximating the absolute performances obtained with real users, exhibiting average differences of 0.015 and 0.107, respectively. 

Based on these observations, we conclude that \emph{LLM-US} outperforms \emph{ABUS} in both reproducing the relative ranking and approximating the absolute performance measurements. While \emph{LLM-US} generally aligns with user satisfaction in ranking CRSs sharing an underlying architecture, it struggles with a more diverse set of systems. This suggests that user simulation, particularly \emph{LLM-US}, presents a viable approach for automating the evaluation of different versions of the same CRS but requires further development to reliably assess broader range of CRSs.

\begin{table}
    \centering
    \caption{Evaluation of CRSs in terms of Reward-per-Dialogue-Length with simulated users. Absolute score differences between measurements obtained with real users and simulated users is shown in parentheses.}
    \label{tab:utility-diff}
    \vspace*{-0.7\baselineskip}
    \begin{tabular}{lcc}
        \toprule
        \textbf{CRS} & \textbf{ABUS} & \textbf{LLM-US} \\
        \midrule
        ChatCRS\_OpenDialKG & -- & 0.074 \footnotesize{(0.037)} \\
        ChatCRS\_ReDial & 0.121 \footnotesize{(0.096)} & 0.049 \footnotesize{(0.024)} \\
        BARCOR\_ReDial & 0.126 \footnotesize{(0.102)} & 0.009 \footnotesize{(0.015)} \\
        BARCOR\_OpenDialKG & 0.117 \footnotesize{(0.099)} & 0.011 \footnotesize{(0.007)} \\
        UniCRS\_ReDial & 0.079 \footnotesize{(0.060)} & 0.026 \footnotesize{(0.007)} \\
        KBRD\_ReDial & 0.084 \footnotesize{(0.076)} & 0.016 (0.008) \\
        CRB-CRS\_ReDial & 0.070 \footnotesize{(0.047)} & 0.011 \footnotesize{(0.012)} \\
        UniCRS\_OpenDialKG & 0.091 \footnotesize{(0.085)} & 0.028 \footnotesize{(0.022)} \\
        KBRD\_OpenDialKG & 0.298 \footnotesize{(0.294)} & 0.000 \footnotesize{(0.004)} \\
        \midrule
        Average abs. diff. & 0.107 & 0.015 \\ 
        \bottomrule
    \end{tabular}
\end{table}

\section{Discussion and Conclusion}

In this work, we addressed the limitations of traditional CRS evaluation by proposing a novel reward-cost framework for measuring user utility and demonstrating its alignment with user satisfaction through two new metrics. Furthermore, we explored the potential of user simulation for automated CRS evaluation, comparing the reliability of different simulation approaches.
We now conclude by discussing several crucial aspects for future work in the automatic evaluation of CRSs. These considerations are informed by our examination of current automatic evaluation practices and simulation-based evaluation of CRSs as a solution to alleviate the dependence on real users and human annotators.\\

\noindent
\emph{\textbf{Choice of evaluation metrics.}} Different metrics provide different insights and can lead to different conclusions on performance, as our experiments illustrate. We acknowledge that the evaluation of CRSs is still an open and complex challenge, hence, an agreed-upon set of such metrics has not been established yet. We advocate for researchers to prioritize metrics that effectively proxy \emph{utility} for the user as the ultimate goal.
To foster the development in this direction, we have presented a general Reward/Cost framework and shown that existing evaluation metrics (Recall@N and success rate) can be understood as specific instantiations within this framework. We encourage future work to explore novel metrics through the lens of user-perceived reward and interaction cost. While we proposed two initial metrics based on single reward and cost factors, showing positive correlation with user satisfaction, future metrics should incorporate a broader range of factors with dynamically adjusted weights to better capture the nuances of user interactions. This need is underscored by direct user feedback from CRSArena-Dial~\citep{Bernard:2025:WSDM}, where some users reported frustration due to, for example, misunderstandings and response time (factors that are not accounted for in our proposed metrics), although the CRS was able to provide satisfactory recommendations.\\

\noindent
\emph{\textbf{Choice of baseline CRSs.}} Our results have indicated a large potential headroom based on self-reported user satisfaction. When benchmarking new CRSs, we recommend that researchers consider at least the following two baselines as representatives of two main classes of approaches: (1) a retrieval-based CRS and (2) a LLM-based CRS. The former is conceptually simpler, requires less resources to train and deploy, and is shown to outperform some of the neural CRSs regarding the proposed Reward-per-Dialogue-Length measure. The latter corresponds to the best type of CRSs with regard to user satisfaction according to our results. \\

\noindent
\emph{\textbf{Reliability of user simulators.}} Our experiments have shown that the LLM-based simulator, \emph{LLM-US}, is more reliable than the agenda-based one, \emph{ABUS}, at evaluating a diverse set of CRSs. However, \emph{LLM-US} still has shortcomings, especially when it comes to accurately reproducing the relative ranking of a diverse set of CRSs. Therefore, we emphasize that results obtained through simulation-based evaluation necessitate validation by human annotators, as our findings highlight that conclusions can vary significantly depending on the specific user simulator employed. \\

\noindent
\emph{\textbf{Reproducibility.}} In practice, a major hurdle to performing simu\-lation-based evaluation is the limited availability and suitability of resources in terms of CRSs and user simulators. To overcome this, it is important to make system implementations publicly available, including a solution for direct interaction (e.g., an API or user interface) and clear setup instructions.
Indeed, when performing this study, we have encountered difficulties implementing direct interactions with CRSs due to unclear descriptions of the data format, required annotations, and parameters expected by the system.
Although some of the CRS authors provided some support, the technical expertise required to operate the system remains high.

\section*{Limitations}

\begin{table}
    \centering
    \caption{Kendall's $\tau$ correlation between utility metrics and self-reported user satisfaction considering a subset of CRSs (\emph{ChatCRS\_*} and \emph{BARCOR\_*}).}
    \label{tab:correlation-metrics-lim}
    \begin{tabular}{l c}
        \toprule
        \textbf{Metric} & \textbf{Kendall's $\tau$} \\
        \midrule
        \textbf{Recall@1} & 0.0 \\
        \textbf{Success rate} & 0.67 \\
        \textbf{Successful rec. round ratio} & -0.34 \\
        \textbf{Reward-per-Dialogue-Length} & 1.0 \\
        \bottomrule
    \end{tabular}
\end{table}

We acknowledge that this work has some limitations.
First, our selection of CRSs and user simulators is not exhaustive and is biased towards publicly available, English-language systems used in previous studies. Our analysis in Section~\ref{sec:expeval:results}, which examines the correlation between the different utility metrics and self-reported user satisfaction, includes all CRSs, even those with very low user satisfaction, which could potentially skew our observations. However, an additional analysis on a subset of the best-performing systems in terms of satisfaction (\emph{ChatCRS\_*} and \emph{BARCOR\_*}) shows that Kendall's $\tau$ correlations with utility metrics remain largely consistent (Table~\ref{tab:correlation-metrics-lim}), with the exception of \emph{SRRR} (which appears to be more sensitive than other metrics). We therefore retained the full set of CRSs to provide a broader perspective on the current state-of-the-art.

Second, our reliance on an LLM for dialogue act annotation may introduce inaccuracies in identifying the acceptance, rejection, and recommendation acts used to compute our proposed utility metrics. We note that the joint annotation of intent and slot values pairs as a multi-label and multi-class classification task is complex and understudied problem~\citep{Weld:2022:CSUR} (proposed models commonly consider a single dialogue act per utterance). For Mistral NeMo, we observed an area under the ROC curve of 0.673 and 0.683 for user and system intents, respectively, across 77 manually annotated dialogues.  While precision scores for \texttt{Accept} (0.61), \texttt{Reject} (0.77), and \texttt{Recommend} (0.96) indicate room for improvement, this approach represents a cost-effective solution for our analysis.

Additionally, some CRSs and user simulators rely on prompts and LLMs that may not be optimal, as different configurations may lead to different results. To mitigate this, the prompts used in this work are inspired and adapted from previous studies~\citep{Terragni:2023:arXiv,Huang:2024:arXiv}. 

From an ethical perspective, the use of LLMs is associated with well-documented concerns, including the potential to generate harmful content, reinforce biases, and violate privacy, in addition to their environmental impact~\citep{Bender:2021:FAccT}. We acknowledge that we did not implement explicit moderation or safeguards to address these important issues.

Finally, the evaluation is conducted solely in the movie domain, focusing on specific aspects of the CRSs, which may limit the generalizability of the findings. Despite this, we believe the proposed methodology is broadly applicable for evaluating CRSs in other domains and languages.

\begin{acks}
This research was supported by the Norwegian Research Center for AI Innovation, NorwAI (Research Council of Norway, project number 309834).
\end{acks}

\balance
\bibliographystyle{ACM-Reference-Format}
\bibliography{sigirap2025-crs-eval.bib}

\end{document}